\input{aipcheck}

\documentclass[
    ,final            
  ]
  {aipproc}

\layoutstyle{6x9}
\newcommand{\aap}{A\&A}

\newcommand{\apj}{ApJ}
\newcommand{\apjl}{ApJL}
\newcommand{\apjs}{ApJS}

\begin{document}

\title{Possible chromospheric activity cycles in II Peg, UX Ari and V711 Tau}

\classification{97.10.Jb,95.85.Mt,97.30.Nr}
\keywords{Stars: activity,binaries close, Ultraviolet: stars,Techniques: spectroscopic}

\author{Andrea P. Buccino}{
}

\author{Pablo J. D. Mauas}{
  address={Instituto de Astronom\'\i a y F\'\i sica del Espacio,
  CONICET-UBA, Buenos Aires, Argentina.} 
}

\begin{abstract}
We study the Mount Wilson indices we obtained indirectly
from IUE high and low resolution spectra of the RS CVn-type
systems II Peg (K2IV), UX Ari (K0IV+G5V) and V711 Tau (K1IV+G5V),
extensively observed by IUE from 1978 to 1996.  We analyze the
activity signatures, which correspond to the primary star, with the
Lomb-Scargle periodogram. From the analysis of V711 Tau data, we found
a possible  chromospheric cycle with a period of  18 years and a shorter
$\sim$3 year cycle, which could be associated to a chromospheric
flip-flop cycle. The data of II Peg also suggest a chromospheric cycle
of $\sim$21 years and a flip-flop cycle of $\sim$9 years.  Finally, we
obtained a possible chromospheric cycle of $\sim$6 years for UX Ari.

\end{abstract}

\maketitle


\section{Introduction}
RS CVn-type stars are binary systems where the most massive primary
component is a G-K giant or subgiant and the secondary is a subgiant or
dwarf of spectral classes G to M. These systems are well known due to
their strong chromospheric plages, coronal X-ray emission, and strong
flares in the optical, UV, radio, and X-ray.

Most long-term stellar activity studies of RS CVn stars are
derived from the easily detected optical photometric variations
produced by their long-lived large spots. In most cases, the mean
magnitude reflects a stellar activity cycle similar to the 11-year
solar one. On the other hand, the peak-to-peak magnitude shows a
shorter cycle, called flip-flop cycle, which reflects the non
axisymmetrical redistribution of the spotted area on the stellar surface.

The IUE database provides a large number of UV high and low resolution
spectra of these type of stars. Furthermore, the IUE satellite
monitored these stars continuously during several seasons. In the
present work, we have measured the Mg II line-core fluxes on the IUE
low and high resolution spectra of three RS CVn-type stars ( II Peg,
UX Ari and V711 Tau) and then converted the fluxes to the Mount Wilson
index $S$
\citep{2006BAAA...49..136B,2008A&A...483..903B}. For each star, we
analyzed the mean annual index $\langle S\rangle$ with the
Lomb-Scargle periodogram
\cite{1982ApJ...263..835S,1986ApJ...302..757H} to search for a
long-term chromospheric cycle. Following D\'\i az et al. \cite{2007A&A...474..345D},
we obtained the rotational modulation of the $S$-index for several
seasons by fitting the light-curves with a harmonic function. We
analyzed the amplitude of each curve with the Lomb-Scargle periodogram
to search for a chromospheric flip-flop cycle.

\section{II Peg-HD 224085}

\begin{figure}[htb!]
\includegraphics[width=0.65\textwidth]{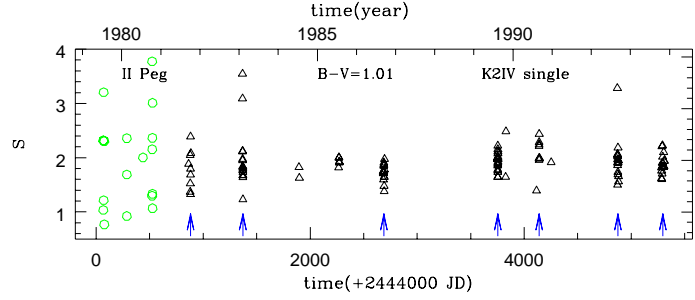}
\caption{II Peg-HD 224085. Mount Wilson index $S$
derived from the IUE high resolution spectra ($\triangle$), and from
the IUE low resolution spectra ($\bigcirc$). Arrows indicate the
seasons for which we analyzed the modulation. }\label{fig.hd224085}
\end{figure}

II Peg (HD 224085) is a single-lined RS CVn-binary system composed by a
K2IV star and an unseen companion of an estimated spectral class M0-M3V
\citep{1998A&A...334..863B}.  
In Fig. \ref{fig.hd224085} we plot the Mount Wilson index $S$ derived
from the IUE high and low resolution spectra obtained between 1979 and
1995.

To search for cyclic patterns in the chromospheric data, we first studied
the mean annual index $\langle S \rangle$ of the data plotted in
Fig. \ref{fig.hd224085} as a function of time with the Lomb-Scargle
periodogram. We obtained a peak at 7741 days ($\sim$21 years) with a
false alarm probability (FAP) of 35\%.

\begin{figure}[htb!]
\centering
\includegraphics[width=0.5\textwidth]{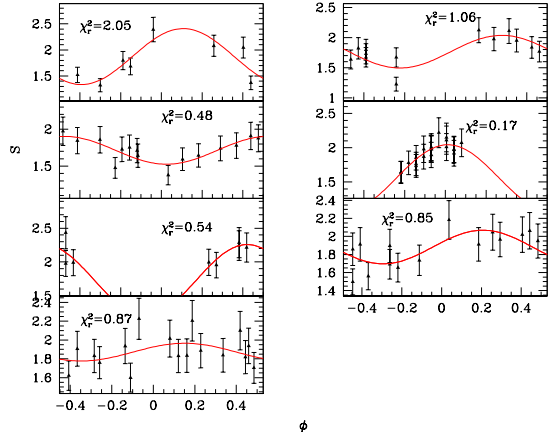}
\small{\caption{Variability on II Peg-HD 220485. Light curves $S$ vs. $\phi$ of the 
    datasets indicated with arrows in Fig. \ref{fig.hd224085}, the
    harmonic curve that best fit the data and the reduced $\chi^2$ of the fit. }\label{fig.hd224085_curvas}}
\end{figure}

Secondly, we  analyzed
the rotational modulation of the index $S$ during the seasons
indicated with arrows in Fig. \ref{fig.hd224085}.
To do so, we phased each
season's light-curve $S$ vs. time with the 6.724-day rotation period \cite{2001A&A...370..974D}
(see Fig. \ref{fig.hd224085_curvas}) and we fitted each set of data
with the harmonic function: $a_0+a_1 \cos(2\pi\phi)+a_2$
\citep{2007A&A...474..345D}.
\vspace{0.2cm}

 We analyzed the amplitude $A=\sqrt{a_1^2+a_2^2}$ against time with the Lomb-Scargle
periodogram and we obtained a cyclic pattern of 3310 $\pm$ 253 days
(9.07 $\pm$ 0.69 years) with a FAP of 31\%. This
periodic behaviour seems to be well correlated with the flip-flop
cycle with a period of 9.30 years obtained by Berdyugina and Tuominen \cite{1998A&A...336L..25B} for
the spot activity.

\section{UX Ari-HD 21242}
\begin{figure}[htb!]
\centering
\includegraphics[width=0.65\textwidth]{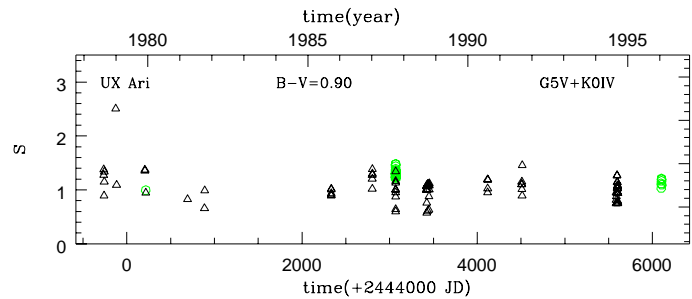}
\caption{UX Ari-HD 21242. Symbols as in Fig. \ref{fig.hd224085}}\label{fig.hd21242}
\end{figure}

UX Ari (HD 21242) is a RS CVn-type system composed by a K0IV star and
a G5V companion in a 6.483-day orbit \citep{2001A&A...370..974D}.
In Fig. \ref{fig.hd21242} we plot the index $S$ obtained for this star
between 1978 and 1996.

We analyzed the mean annual $\langle S\rangle$ of the indexes plotted
in Fig. \ref{fig.hd21242} as a function of time with the Lomb-Scargle
periodogram. We obtained a period of 2180 $\pm$ 32 days ($\sim$6 years) with a
FAP of 29\%.

\section{V711 Tau-HD 22468}
\begin{figure}[htb!]
\centering
\includegraphics[width=0.65\textwidth]{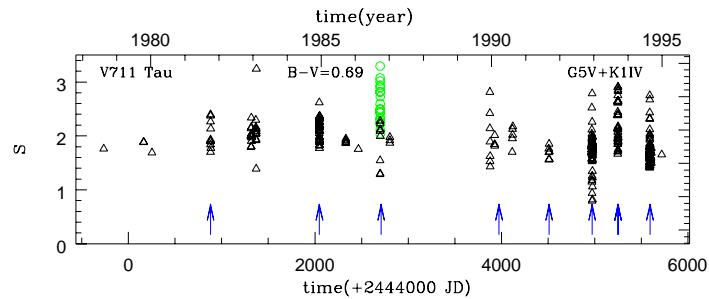}
\caption{V711 Tau-HD 22468. Symbols as in Fig. \ref{fig.hd224085}}\label{fig.hd22468}
\end{figure}

V711 Tau  (HR 1099, HD 22468) is one of the most active RS CVn non-eclipsing
spectroscopic binary system, consisting of a K1 subgiant primary and a G5 dwarf secondary in a
2.837 day-orbit \cite{1983ApJ...268..274F}. 
In Fig. \ref{fig.hd22468} we plot the index $S$  for this star between 1978 and 1995.  

 We analyzed the mean
annual $\langle S\rangle$  as a function of time with the Lomb-Scargle
periodogram and we obtained a peak at 6589$\pm$1170 days  with
a FAP of 11\%. This period of $18.05\pm
3.21$ years is in agreement with the ones reported in the literature
\citep{1995ApJS...97..513H,2006A&A...455..595L,2007ApJ...659L.157B}.

\begin{figure}[htb!]
\centering
\includegraphics[width=0.5\textwidth]{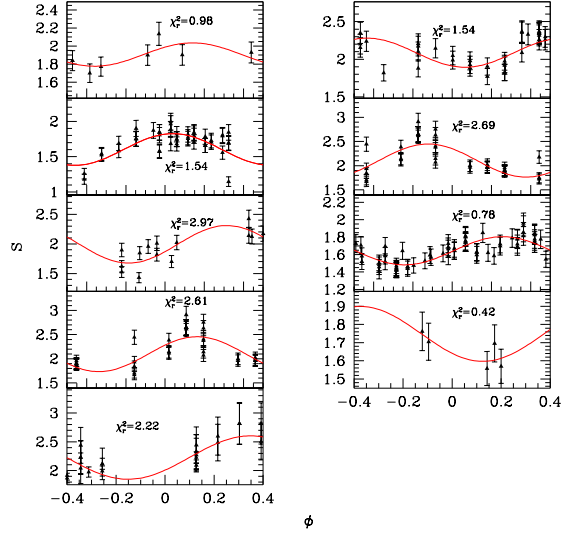}
\small{\caption{Variability on  V711 Tau. Light curves $S$ vs. $\phi$ of the 
    datasets indicated with arrows in Fig. \ref{fig.hd22468}, the
    harmonic curve that best fit the data and the reduced $\chi^2$ of
    the fit. }\label{fig.hd22468_curvas}}
\end{figure} 

On the other hand, we analyzed the amplitude of the rotational modulation
of the datasets indicated with arrows in Fig. \ref{fig.hd22468} and we
obtained a cyclic pattern in the amplitude $A$ of the curves plotted
in  Fig. \ref{fig.hd22468_curvas} of period 
1207$\pm$ 45 days ($\sim$ 3.3 years) of 23\% FAP, which is consistent
with the one obtained by Lanza et al. \cite{2006A&A...455..595L} for
the spot activity.

\begin{theacknowledgments}We thank the economical support provided by
the Project 05-32408 of the Agencia de Promoci\'on Cient\'\i fica y
Tecnol\'ogica and by the ``Cool Star 15'' organising
 commitee. \end{theacknowledgments}



\bibliographystyle{aipproc}   



\end{document}